\documentclass[a4paper]{llncs}

\usepackage{times}
\usepackage{helvet}
\usepackage{courier}
\usepackage{graphicx}
\usepackage{subfigure}
\graphicspath{{./images/}}

\begin{document}

\title{Detection and Analysis of 2016 US Presidential Election Related Rumors on Twitter}

\author{Zhiwei Jin$^{1,2}$, Juan Cao$^{1,2}$, Han Guo$^{1,2}$, Yongdong Zhang$^{1,2}$, Yu Wang$^{3}$ \and Jiebo Luo$^{3}$
}
\institute{
   $^{1}$Key Laboratory of Intelligent Information Processing,\\
    Institute of Computing Technology, CAS, Beijing 100190, China\\
   $^{2}$University of Chinese Academy of Sciences, Beijing 100049, China\\
   $^{3}$University of Rochester, Rochester, NY 14627, USA\\
   \{jinzhiwei, caojuan, guohan, zhyd \}@ict.ac.cn; ywang.tsinghua@gmail.com; jluo@cs.rochester.edu
}
\maketitle
\begin{abstract}
The 2016 U.S. presidential election has witnessed the major role of Twitter in the year's most important political event. Candidates used this social media platform extensively for online campaigns. Meanwhile, social media has been filled with rumors, which might have had huge impacts on voters' decisions. In this paper, we present a thorough analysis of rumor tweets from the followers of two presidential candidates: Hillary Clinton and Donald Trump. To overcome the difficulty of labeling a large amount of tweets as training data, we detect rumor tweets by matching them with verified rumor articles. We analyze over 8 million tweets collected from the followers of the two candidates. Our results provide answers to several primary concerns about rumors in this election, including: which side of the followers posted the most rumors, who posted these rumors, what rumors they posted, and when they posted these rumors. The insights of this paper can help us understand the online rumor behaviors in American politics.
\end{abstract}

\section{Introduction}

In the 2016 U.S. presidential election, Twitter became a primary battle ground: candidates and their supporters were actively involved to do campaigns and express their opinions by tweeting \cite{wang2016catching}. Meanwhile, the fact that various rumors were spreading on social media during the election became a serious concern. Among all the 1,723 checked rumors from the popular rumor debunking website Snopes.com, 303 rumors are about Donald Trump and 226 rumors are about Hillary Clinton. These rumors could potentially have negative impacts on their campaigns.

\begin{figure}
\centering
\includegraphics[width=3.5in]{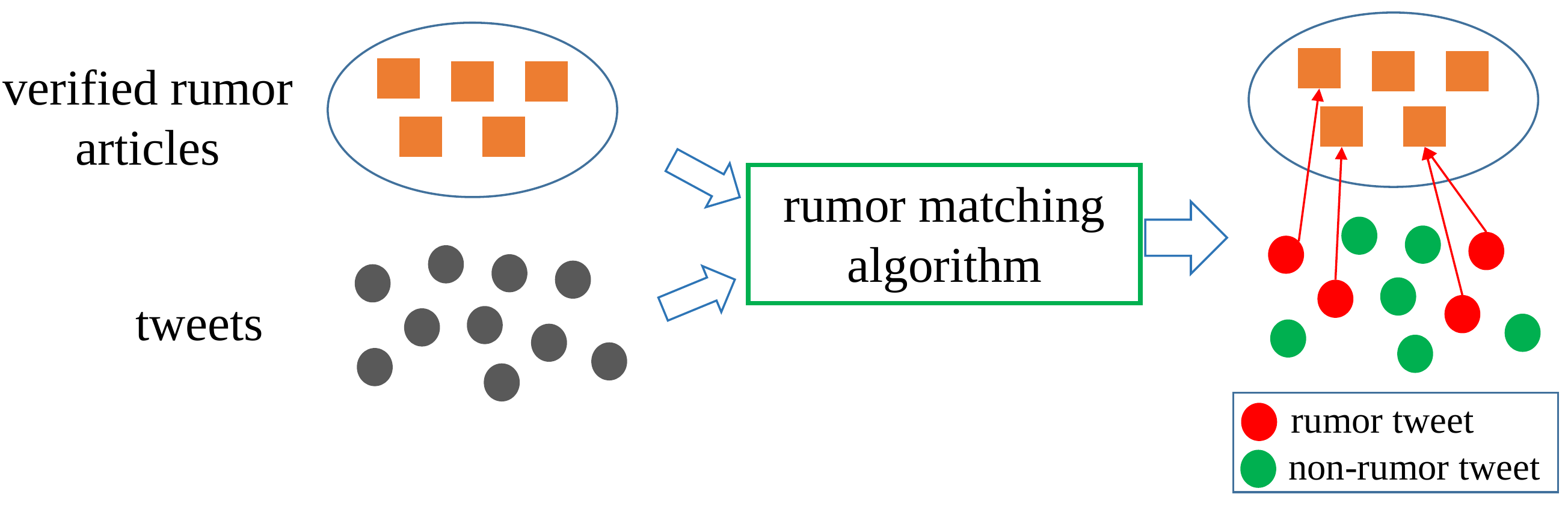}
\caption{Rumor detection as a text matching task.}
\label{fig_3}
\end{figure}

In this paper, we aim to understand the rumor spreading behaviors of candidates' followers. A rumor is defined as a controversial and fact-checkable statement \cite{difonzo2007rumor}. Existing machine learning methods for rumor detection \cite{castillo2011information}\cite{jin2016}\cite{wu2015false} commonly require extensive labeled training data, which is expensive to label for the rumor detection problem. Besides, it is difficult to tell what rumors are posted as their binary results are not easily interpretable. Considering these limitations, we use the checked rumors from Snope.com as the objective golden samples and propose to detect rumors as a text matching task (Fig. \ref{fig_3}). In this scheme, a set of verified rumor articles are collected as standard samples for reference. Each tweet is compared with these verified rumors to see if they match closely. Compared with existing approaches, our approach requires minimal human labeling and the matching results can be easily interpreted.

In order to find the best matching algorithm, we conduct a comparative study of several competing algorithms. These algorithms are executed on a reasonably sized set of 5,000 manually labeled tweets to provide a fair performance comparison. We then detect rumors with the selected most effective matching algorithm on over 8 million of tweets from 14,000 followers of the two leading presidential candidates: Hillary Clinton and Donald Trump. We inspect the rumor detection results from different aspects to answer following questions: which side posted the most rumors? who posted these rumors? what rumors did they post? when did they post rumors? These insights help us understand the rumor tweeting behaviors of different groups of followers and can be helpful for mining voters' real intentions and accurately detecting rumors during political events in the future.

\section{Related Work}

Online social media have gained huge popularity around the world and become a vital platform for politics. However, the openness and convenience of social media also fosters a large amount of fake news and rumors which can spread wildly \cite{friggeri2014rumor}. Compared with existing rumor detection works that are focused on general social events or emergency events \cite{jin2014news}, this paper presents a first analysis of rumors in a political election.

Most existing rumor detection algorithms follow the traditional supervised machine learning scheme. Features from text content \cite{castillo2011information}, users, propagation patterns \cite{wu2015false} and multimedia content \cite{jin2015,jin2016novel} are extracted to train a classifier on labeled training data. Some recent works further improve the classification result with graph-based optimization methods \cite{gupta2012evaluating,jin2014news,jin2016}.

Although machine learning approaches are very effective under some circumstances, they also have drawbacks. The supervised learning process requires a large amount of labeled training data which are expensive to obtain for the rumor detection problem. They derive features in a ``black box" and the classification results are difficult to interpret. In \cite{zhao2015enquiring}, a lexicon-based method is proposed for detecting rumors in a huge tweet stream. They extracted some words and phrases, like ``rumor", ``is it true", ``unconfirmed", for matching rumor tweets. Their lexicon is relatively small, thus the detection results tend to have high precision but low recall of rumors.

In this paper, we formulate the rumor detection as a text matching task. Several state-of-the-art matching algorithms are utilized for rumor detection. TF-IDF \cite{sparck1972statistical} is the most commonly used method for computing documents similarity. BM25 algorithm \cite{robertson2009probabilistic} is also a term-based matching method. Recent research in deep learning for text representation embeds words or documents into a common vector space. Word2Vec \cite{mikolov2013distributed} and Doc2Vec \cite{le2014distributed} are two widely used embedding models at the word and paragraph levels, respectively.

\section{Dataset}
We collect a large-scale dataset for analyzing rumors during the 2016 U.S. presidential election from Twitter. For reliable rumor detection, we obtain a set of verified rumor articles from Snopes.com. We also manually construct a testing set to fairly evaluate the rumor detection methods.

Using the Twitter API, we collect all the users who are following the Democratic presidential candidate Hillary Clinton and the Republic presidential candidate Donald Trump. We randomly select about 10,000 followers from each candidate's follower list, which contains millions of followers. We then collect up to 3,000 most recent tweets for each user using the Twitter API. Altogether, we get 4,452,087 tweets from 7,283 followers of Clinton and 4,279,050 tweets from 7,339 followers of Trump in our dataset.

We collect a set of verified rumor articles from Snopes.com as gold standard samples for rumor matching. Snopes.com is a very popular rumor debunking website. Social media users can nominate any potential rumor to this site. The employed analysts then select some of these controversial statements to fact-check them as rumors or truth. An article is presented for each checked rumor by these professional analysts, which gives conclusion of the rumor followed by full description, source, origin, suporting/opposing evidences of the rumor story. We collect the articles of all the 1,723 checked rumors on this website to form the verified rumor article set.

To quantitatively evaluate the performance of rumor detection methods, we build a manually labeled tweet set. We randomly select 100 rumors from the verified rumor set. For each verified rumor article, we search the large tweet set with keywords extracted from the article. Each tweet in the search result is manually examined to check if it matches the rumor article. After these procedures, we obtain a set of 2,500 rumor tweets from 86 rumor articles. We then randomly sample the same number of unrelated tweets as negative samples. In this set, not only is each tweet labeled as rumor or not, but the rumor tweets are also labeled with their corresponding verified rumor articles. Therefore, we can perform both general rumor classification and fine-grained rumor identification with this dataset. The following is an example of a verified rumor article and three associated tweets.

\textbf{Verified rumor article}\footnote{The full article is available at: http://www.snopes.com/hillary-clinton-has-parkinsons-disease/}:

{\it Shaky Diagnosis. A montage of photos and video clips of Democratic presidential candidate Hillary Clinton purportedly demonstrates she has symptoms of Parkinson's disease. Photos and video clips narrated by a medical doctor demonstrate that Democratic presidential candidate Hillary Clinton likely has Parkinson's disease......}

\textbf{Associated rumor tweets}:
\begin{enumerate}
  \item {\it Hillary collapse at ground zero! game over, Clinton! Parkinson's blackout!}
  \item {\it Wikileaks E-mails: Hillary looked into Parkinson's drug after suffering from ``decision fatigue".}
  \item {\it Exclusive Report: How true is this?? Hillary Clinton has Parkinson's disease, doctor confirms.}
\end{enumerate}

\section{Rumor Detection}
We formulate rumor detection on Twitter as a matching task in this paper (Fig. \ref{fig_3}). With reliable rumor articles collected from Snopes.com, the key part of this scheme is the matching algorithm. Compared with the traditional rumor classification algorithms, our rumor matching scheme not only outputs a tweet as rumor or not but also identifies which rumor article it refers to if it is a rumor tweet. We perform comparative studies of different matching algorithms on both the classification and the identification task of rumor detection.

\subsection{Rumor Detection Algorithms}

We compare the performance of five matching algorithms with respect to the rumor detection task. The first set of methods includes two widely used term-based matching methods: TF-IDF and BM25. The second set includes two recent semantic embedding algorithms: Word2Vec and Doc2Vec. The third set is a lexicon-based algorithm for rumor detection on Twitter stream.

\textbf{TF-IDF} \cite{sparck1972statistical} is a widely used model in text matching. In this model, both the tweets and the verified rumor articles are represented as a $v$-dimensional vector, where $v$ is the size of the dictionary of the corpus. Each element in the vector stands for the TF-IDF score of the corresponding word in the text. TF is the term frequency. IDF score is the inverse document frequency, which is calculated on the whole corpus.

\textbf{BM25} \cite{robertson2009probabilistic} is also a text similarity computing algorithm based on the bag-of-words language model. It is an improvement of the basic TF-IDF model by normalizing on term frequency and document length. Both TF-IDF and BM25 have been widely used in many related studies.

\textbf{Word2Vec} \cite{mikolov2013distributed} represents each word in a corpus with a real-valued vector in a common semantic vector space. Compared with traditional lexical-based matching models, this algorithm evaluates the quality of word representations based on their semantic analogies. We use the pre-trained Word2Vec model on a corpus of 27 billion tweets. The word dimension is 200. To aggregate a presentation for a whole text, we take the average of word vectors in the text.

\textbf{Doc2Vec}  \cite{le2014distributed} is also an embedding algorithm on the semantic space, which can directly learn the distributed representations of documents. We use all the tweets and rumor articles for the unsupervised training of the model after standard pre-processing. We use the default parameter settings as in \cite{le2014distributed}. After training, tweets and verified rumors are represented as 400-dimensional vectors.

For Word2Vec and Doc2Vec, the matching score between a tweet and a rumor article is computed based on the cosine distance of their vector representions.

\textbf{Lexicon matching} \cite{zhao2015enquiring} is a lexicon-based rumor detection algorithms for efficiently detecting in huge tweet streams. It mines a couple of signal words or phrases for recognizing prominent rumor tweets. We use the same set of regular expression patterns as in \cite{zhao2015enquiring} to match rumor tweets.

\subsection{Evaluation on Rumor Classification Task}

\begin{figure}
\centering
\includegraphics[width=2.5in]{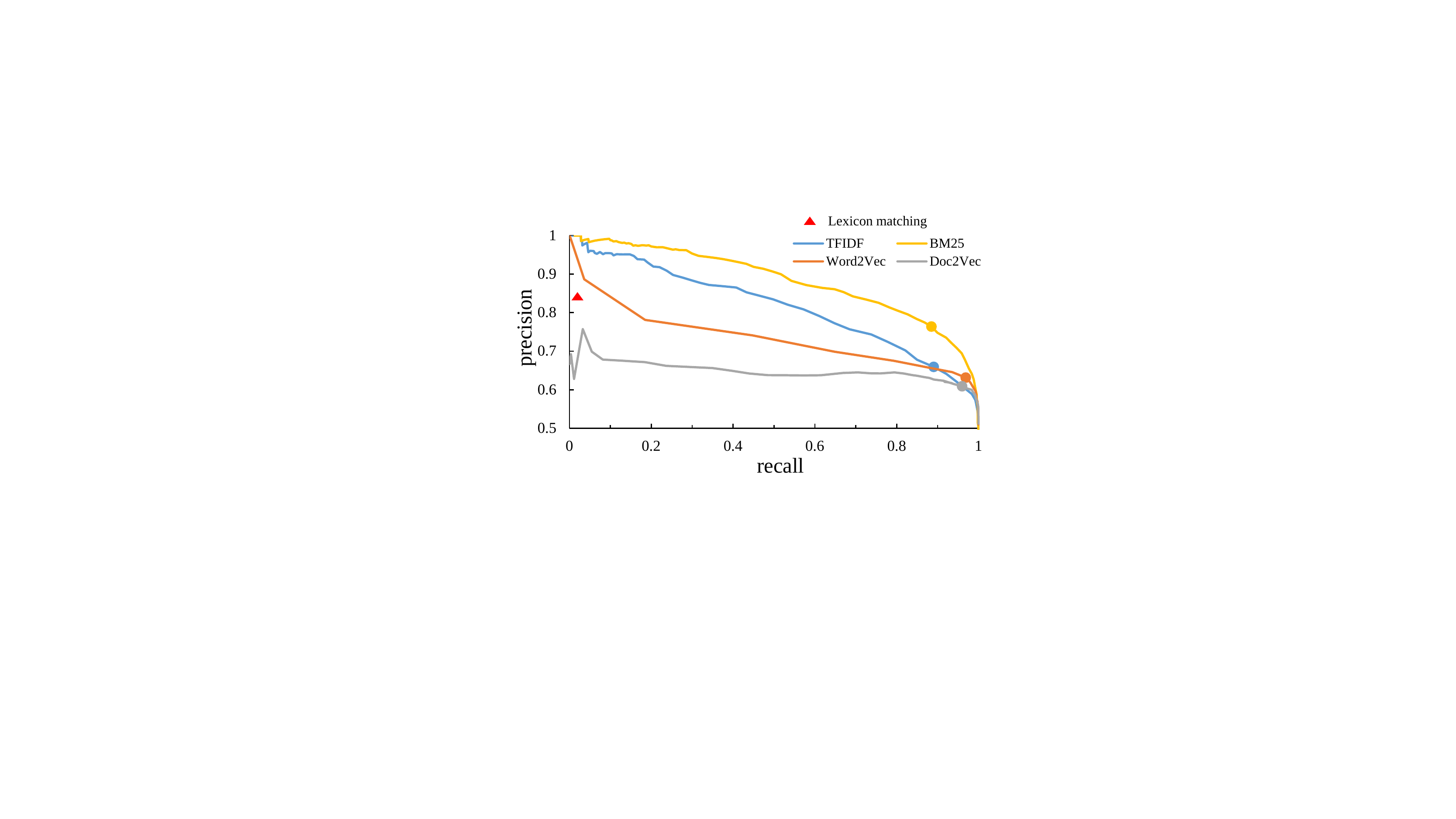}
\caption{The comparative performance of four matching algorithms.}
\label{fig_4}
\end{figure}

TF-IDF, BM25, Word2Vec and Doc2Vec represent texts as numeric vectors. The similarity between a tweet and a verified rumor is computed as their matching score. By setting a threshold $h$ for each method, we classify tweets with matching scores larger than $h$ as rumor tweets. We can achieve different precision and recall of rumor classification by varying the threshold. We test all the four methods on the 5,000 labeled tweet set. Fig. \ref{fig_4} illustrates the precision-recall curves of these four algorithms. The lexicon matching algorithm detects rumors by keywords matching, thus its result is actually fixed (as a single point in Fig. \ref{fig_4}).

The highlighted round points on each curve in Fig. \ref{fig_4} are points where the F1-measures are maximized, at 0.758, 0.82, 0.764 and 0.745 for TF-IDF, BM25, Word2Vec and Doc2Vec, respectively. The red triangle is the fixed result of lexicon matching. These results show that BM25 reaches the best performance among all the five rumor classification methods under different metrics. The two term-based methods (TF-IDF and BM25) outperform the semantic-embedding and lexicon-based methods. For semantic-embedding, Word2Vec is slightly better than Doc2Vec. Lexicon matching can reach a rumor classification precision of 0.862, but its recall (0.008) is too low.

\subsection{Evaluation on Rumor Identification Task}
One extra advantage of our proposed rumor matching scheme is its ability to identify what rumor article a rumor tweet refers to, apart from classifying it as a rumor tweet. To compare the rumor identification performance of the four algorithms, we compute the similarity score between each pair of tweet and verified rumor article for the 2,500 labeled rumor tweets and 1,723 verified rumor articles. If the most similar rumor article of a tweet is exactly the same labeled rumor article for it, then this is an accurate rumor identification.

\begin{table}
\centering
\caption{The accuracy of rumor identification task.}
\label{table_1}
\begin{tabular}{c|c|c|c|c}
\hline
& TF-IDF &  BM25 & Word2Vec & Doc2Vec  \\
\hline
Accuracy & 0.795 & \textbf{0.799} & 0.557 & 0.658\\
\hline
\end{tabular}
\end{table}

From the overall rumor identification accuracy of each rumor matching methods, the BM25 algorithm achieves the best accuracy of 0.799. The accuracy of BM25 is only slightly better than that of TF-IDF, although it has major advantage in the rumor classification task. This is probably because BM25 can distinguish non-rumor tweets much better than TF-IDF. Another interesting finding is that Doc2Vec actually performs better on the rumor identification task than Word2Vec, although the latter has slightly better performance on the rumor classification task.

\section{Analyzing Rumor Tweets Pertaining to the Election}
This paper analyzes rumor tweets related to the 2016 U.S. presidential election. For rumor analysis at a large scale, in this section, we use the proposed rumor detection algorithm to detect rumor tweets from over 8 million tweets collected from the followers of Hillary Clinton and Donald Trump. Specifically, we match each rumor tweet with corresponding rumor articles in the verified set with BM25 algorithm. To conduct a reliable and accurate analysis, we prefer a high precision for our rumor detection result. We set the similarity threshold $h=30.5$ so that we can achieve a very high rumor classification precision of 94.7\% and the recall of 31.5\% on the test set. Based on the results, we obtain insights into the rumor tweeting behaviors from various aspects.

\subsection{Which side posted the most rumors?}
Twitter became an online battle field during the election. The number of rumor tweets reflects the involvement of candidates' followers in the election campaign. Which side of followers were involved most in spreading rumor tweets? To answer this question, we use rumor classification method to detect rumors in the subset of tweets of the two candidates, respectively. Given our focus on rumors during the election period, we also analyze rumor tweets posted from April up to the present.

\begin{table}
\centering
\caption{Rumor tweet ratio of two candidate's follower groups.}
\label{table_2}
\begin{tabular}{c|c|c}
\hline
& Clinton's followers &  Trump's followers  \\
\hline
Entire time & 1.20\% & 1.16\% \\
\hline
Election period & 1.26\% & 1.35\% \\
\hline
\end{tabular}
\end{table}

From the results in Table \ref{table_2}, we find that:
\begin{itemize}
  \item For entire time, Clinton's followers are slightly more active in posting rumor tweets than Trump's followers. 1.2\% tweets are rumor tweets from Clinton's followers, which is about 4\% more than that of Trump's followers.
  \item People tend to post more rumor tweets in the election time than in the whole time, especially for Trump's followers. Comparing their election period and all time rumor tweeting, Trump's followers have a rumor tweet ratio of 1.35\% during the election, which is 18\% higher than that in all time.
  \item During the election time, Trump's followers are more active in rumor tweeting than Hillary's followers. As the figure suggests, Trump's followers become much more involved in posting rumors at the election time, compared with Clinton's followers.
\end{itemize}

\subsection{Who posted these rumors?}

Who are behind the rumors spreading on Twitter? We investigate this issue by analyzing rumor tweets posted by individual followers of the two candidates.

We rank users by the total number of rumor tweets they posted. We find that the majority of rumors are posted by only a few users: for both Trump's and Clinton's followers, the top 10\% users posted about 50\% rumor tweets and the top 20\% users posted about 70\% of all rumor tweets.

\begin{figure}
\centering
\includegraphics[width=2.5in]{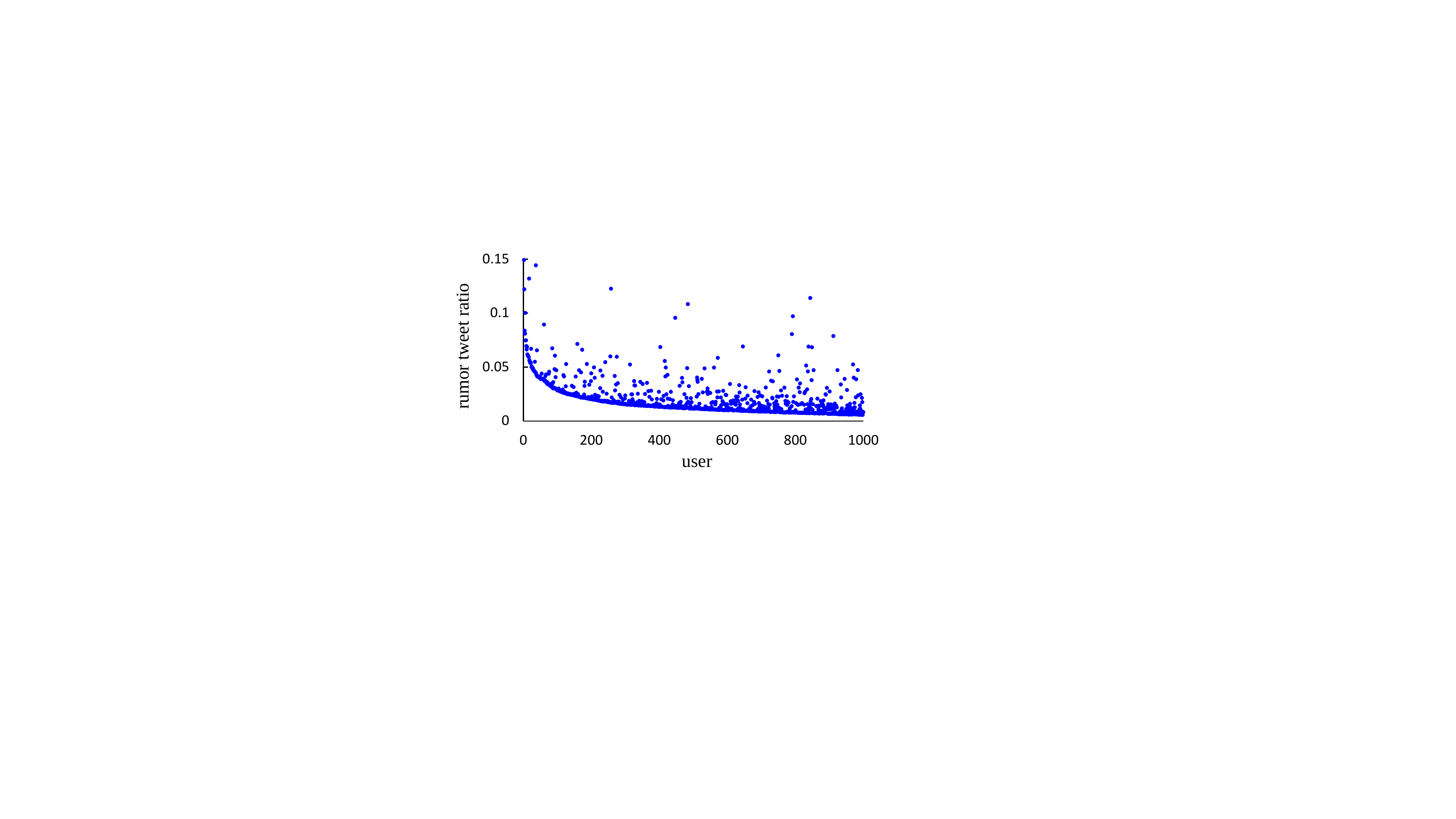}
\caption{Rumor tweet ratio of Clinton's followers.}
\label{fig_8}
\end{figure}

Are these rumor-prolific followers just active in tweeting rumors or active in general tweeting as well? To understand this, we calculate the ratio of rumor tweets in all tweets posted by a user. We rank users based on the rumor ratio in their tweets. In Fig. \ref{fig_8}, we show the top 1000 users from Clinton's followers. We observe that followers who post more rumor tweets also tend to have a larger rumor tweet ratio. This means the rumor-prolific users did not randomly post any tweets; they were actually more concentrated on posting rumor tweets than the users who occasionally post a few rumor tweets.

\textbf{Case Study} After analyzing rumor spreaders at a large scale, we can also conduct a detailed analysis for a specific user.

Take one of Trump's followers, for example. This user posted 3,211 tweets in our dataset, 307 of which are detected as rumors. The rumor tweet ratio is as high as 9.6\%, which means this user is very active in rumor tweeting. By examining the top keywords in all tweets posted by the user (Table \ref{table_3}), we find this person is very focused on posting tweets about the 2016 presidential election: ``Clinton'', ``Sanders'', ``Trump'' and ``election'' are the most mentioned words in the tweets. After rumor detection, we find that the rumor tweets of this user are mainly about Clinton and Sanders rather than Trump: 15\% tweets about Clinton and 28\% tweets about Sanders are rumor tweets, while only 10\% tweets about Trump are rumors.

\begin{table}
\centering
\caption{The number of keywords in the tweets posted by one follower of Trump.}
\label{table_3}
\begin{tabular}{c|c|c|c|c|c|c}
\hline
 & Clinton & Sanders & Trump & election & Democratic & FBI  \\
\hline
Rumor & 1,106 & 620 & 275 & 271 & 97 & 88 \\
\hline
Nonrumor & 197 & 247 & 34 & 30 & 54 & 15 \\
\hline
\end{tabular}
\end{table}

\subsection{What rumors did they post?}
During the election, most rumors are focused on the candidates. By analyzing what people from different groups tweeted about in rumors, we can understand their intentions in this election. We use BM25 to identify the content of each rumor tweet by matching it with the verified rumor articles from Snopes.com. Given our focus on the two primary presidential candidates, Hillary Clinton and Donald Trump, we only analyze rumor tweets related to them. After normalizing the number of candidate-related rumor tweets with the total number of rumor articles for this candidate in our dataset, we plot the rumor content spread by Trump's and Clinton's followers in Table \ref{table_4}. We offer some analysis of this figure based on the normalized rumor tweet number.

\begin{table}
\centering
\caption{Normalized number of rumors posted by followers of Trump and Clinton.}
\label{table_4}
\begin{tabular}{l|c|c}
\hline
& Clinton's followers &  Trump's followers  \\
\hline
Rumors about Clinton & 50.31 & 54.50 \\
\hline
Rumors about Trump & 53.95 & 51.94 \\
\hline
\end{tabular}
\end{table}

First, both follower groups post rumors about their favored candidate as well as the opponent candidate. Supporters of one candidate would spread rumors about the opponent as a negative campaign tactic and debunk rumors about their favored candidate. For example, we show two tweets about the rumor ``{\it Hillary Clinton has Parkinson's disease}'' from our dataset:

\textbf{Tweet 1}: {\it Medical experts watching debate said Hillary showed ``Telltale Signs'' of Parkinson's Disease.}

\textbf{Tweet 2}: {\it ``I know her physician; I know some of her health history which is really not so good'' Trump's MD on Hillary---her MD shared her info with him?}

The first tweet comes from a follower of Trump. It is spreading the rumor by quoting medical experts. The second tweet comes from a follower of Clinton. It is questioning the truthfulness of the rumor.

Second, users would post more rumor tweets about the opponent candidate than their favored candidate. Clinton's followers post 8\% more rumor tweets about Trump than rumors about Clinton. Trump's followers post 5\% more rumor tweets about Clinton than rumors about Trump. Moreover, Trump's followers are more active in this rumor tweeting behavior towards both Clinton and Trump. The numbers of rumor tweets about the two candidates posted by Trump's followers are both larger than those of Clinton's followers.

\subsection{When did they post these rumors?}

\begin{figure}
\centering
\includegraphics[width=3.5in]{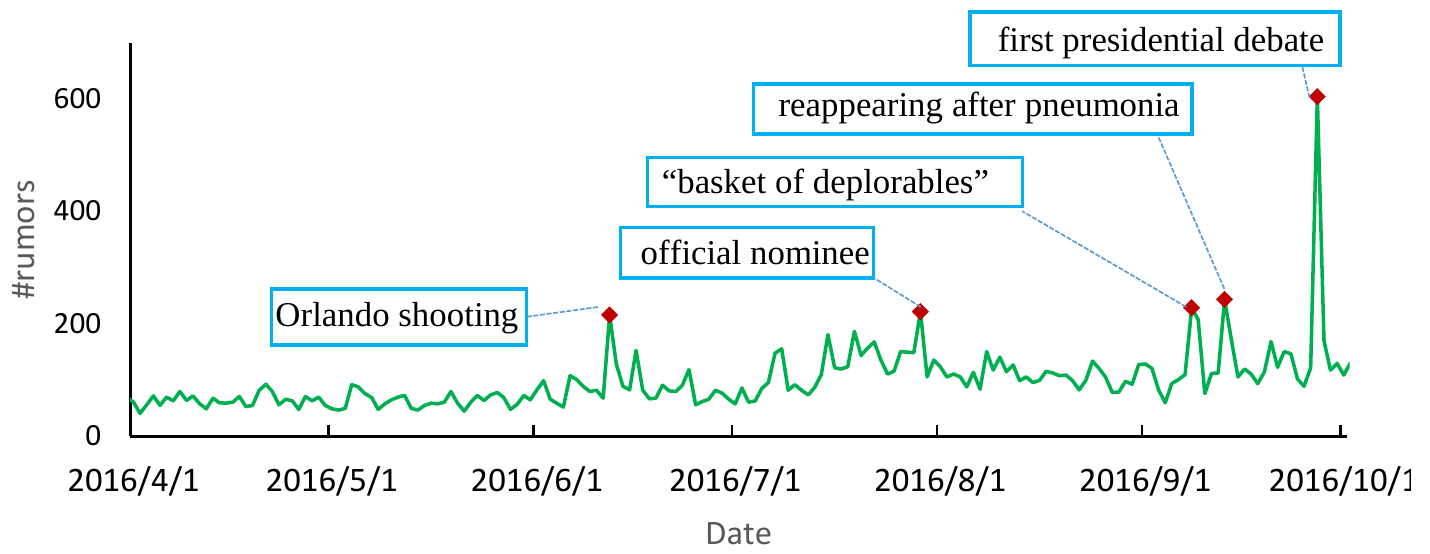}
\caption{Rumor tweet timeline of Clinton's followers.}
\label{fig_11}
\end{figure}

Analyzing the time patterns of rumor tweeting can reveal insights of online campaign. We plot the rumor tweeting of Clinton's followers over six months (April 2016 to September 2016) in Fig. \ref{fig_11}. We annotate the key events for some rumor peaks in the figure to understand the inherent reason behind them. We find that rumors are peaked in three types of occasions: 1) key point in the presidential campaign, such as ``the presidential debate'' and ``official nominee''; 2) controversial emergency events, including ``the Orlando shooting''; 3) events triggering rumors, such as ``reappearing after pneumonia''. This insight reminds us to pay more attention to rumors during these types of events in future political campaigns.

\section{Conclusions}
This paper studies the rumors spreading phenomenon on Twitter during the 2016 U.S presidential election. We propose a reliable and interpretable approach to detecting rumor tweets by matching them with verified rumor articles. We conduct a comparative study of five algorithms for this rumor matching approach. With a rumor detection precision of 94.7\%, we use this method to detect rumors in over eight million tweets collected from the followers of the two primary presidential candidates. We provide a thorough analysis on the detected rumor tweets from the aspects of people, content and time. We would benefit from the discovery in the paper to understand rumors during political events and build more effective rumor detection algorithms in the future.

\subsubsection*{Acknowledgments.}
This work was supported in part by the National Key Research and Development Program of China under Grant 2016YFB0800403 and the National Nature Science Foundation of China (61571424, 61525206). Jiebo Luo and Yu Wang would like to thank the support from the New York State through the Goergen Institute for Data Science. Zhiwei Jin gratefully thanks the sponsorship from the China Scholarship Council.

\bibliographystyle{splncs03}
\bibliography {ref}
\end{document}